\documentclass[onecolumn]{aastex6}
\usepackage{amsmath}
\usepackage{amsfonts}
\usepackage{amssymb}
\usepackage{multirow}
\usepackage{graphicx}
\usepackage{booktabs}

\newcommand{\adsurl}[1]{\href{#1}{ADS}} 
\providecommand{\url}[1]{\href{#1}{#1}}

\def \apjs {{\rm {ApJS}}}

\def \aap {{\rm {A\&A}}}

\def \apjl{\rm {ApJL}}

\begin{document}
\title{`Oumuamua as a messenger from the Local Association}
\author{F. Feng\altaffilmark{1}, H. R. A. Jones}
\affil{Centre for Astrophysics Research, School of Physics, Astronomy and Mathematics, University of Hertfordshire, College Lane, Hatfield AL10 9AB, UK}

\date{\today}

\altaffiltext{1}{fengfabo@gmail or f.feng@herts.ac.uk}

\begin{abstract}
  With a hyperbolic trajectory around the Sun, `Oumuamua is the first confirmed interstellar object. However, its origin is poorly known. By simulating the orbits of 0.23 million local stars, we find 109 encounters with periastron less than 5\,pc. `Oumuamua's low peculiar velocity is suggestive of its origin from a young stellar association with similar velocity. In particular, we find that 'Oumuamua would have had slow encounters with at least five young stars belonging to the Local Association thus suggesting these as plausible sites for formation and ejection.
In addition to an extremely elongated shape, the available observational data for ‘Oumuamua indicates a red colour suggestive of a potentially organic-rich and activity-free surface. These characteristics seem consistent with formation through energetic collisions between planets and debris objects in the middle part of a young stellar system. We estimate an abundance of at least 6.0$\times10^{-3}$\,au$^{-3}$ for such interstellar objects with mean diameter larger than 100\,m and find that it is likely that most of them will be ejected into the Galactic halo. Our Bayesian analysis of the available light curves indicates a rotation period of $6.96_{-0.39}^{+1.45}$\,h which is consistent with the estimation by \cite{meech17} and shorter than other literature. The codes and results are available on \href{https://github.com/phillippro/Oumuamua}{GitHub}.
\end{abstract} 
\section{Introduction}     \label{sec:introduction}
A/2017 U1 was discovered in the Pan-STARRS survey in Hawaii by Robert Weryk and was later found to be on a hyperbolic orbit with an eccentricity of $e=1.1994\pm0.0002$, semi-major axis of $a=-1.2805\pm0.0009$, perihelion of $q=0.25529\pm7.8\times10^{-5}$, and an inclination of $i=122.682\pm0.007$ based on the NASA/JPL Horizons
      On-Line Ephemeris System \citep{giorgini01}. This leads to a pre-encounter velocity of 26.33$\pm$0.01\,km/s and $(U, V, W)=(-11.427\pm0.006, -22.425\pm0.004, -7.728\pm0.007)$\,km/s. Its high pre-encounter velocity strongly favor an interstellar origin leading to
its Hawaiain name of Oumuamua is intended to reflect that this object is like a scout or messenger sent from the past to reach out to us. As the first interstellar object discovered in the Solar System its original discovery name of A/2017 U1 has been revised by the International Astronomical Union with the new designation of ``I'' for interstellar objects, with ʻOumuamua being designated 1I and maybe known as 1I, 1I/2017 U1, 1I/`Oumuamua and 1I/2017 U1 (`Oumuamua)\footnote{\url{https://www.minorplanetcenter.net/mpec/K17/K17UI1.html}}.

No coma of `Oumuamua is detected \citep{jewitt17} and spectroscopic observations do not show any signs of activity on `Oumuamua \citep{fitzsimmons17, masiero17,ye17,meech17}. Thus it is probably an asteroid ejected from the warm part of an extra-solar system \citep{ye17}. Photometric monitoring of this target support a double peaked rotation period of about 8\,h \citep{knight17,bolin17,jewitt17} while \citep{meech17} reported a shorter period of 7.34\,h based on more than 100 data points. The rotation-induced magnitude variation suggest a semi-axes of about 230\,m$\times$35\,m, corresponding to a 6:1 axis ratio and thus indicating albedo variation on the surface \citep{jewitt17}. A higher axis ratio is estimated by \cite{meech17}, indicating a rare cigar-shaped body.  

Various scenarios have been proposed to explain `Oumuamua's origin \citep{mamajek17, gaidos17,zwart17}. \cite{gaidos17} argue that it was probably from the young Carina and Columba Associations due to similar UVW velocities. Its small peculiar velocity also suggests a lack of close encounters with stars and thus a short period of drifting in the Galaxy. On the other hand, \cite{mamajek17} and \cite{zwart17} propose the Galactic interstellar-object debris as its origin due to a lack of appropriate candidates for its original home and an apparent thermalization of `Oumuamua's velocity.

In this work, we argue that `Oumuamua was plausibly ejected from a stellar system in the Local Association (or Pleiades moving group; \citealt{montes01}) based on numerical and statistical arguments. The paper is structured as follows. We identify stellar encounters of `Oumuamua based on numerical integration of stellar orbits in section \ref{sec:encounter}. Then we argue that `Oumuamua is young by investigating its kinematics and light curves in section \ref{sec:age}. We discuss and conclude in section \ref{sec:conclusion}. 

\section{Possible locations for the origin of `Oumuamua}\label{sec:encounter}
To find the origin of `Oumuamua, we derive the pre-encounter velocity by integration of the orbit of `Oumuamua backward to AD 1600 using the JPL HORIZONS service, following \cite{mamajek17}. The heliocentric position and velocity of `Oumuamua in the Galactic coordinate system is $(X,Y,Z)=(1011.69\pm0.54, 1982.13\pm0.35, 684.52\pm0.66)$\,au and $(U,V,W)=(-11.427\pm0.006, -22.425\pm0.004, -7.728\pm0.007)$\,km/s, respectively. We then adopt the Galaxy model and the Sun's initial conditions from \cite{feng14} and follow \cite{feng18a} to use the Bulirsch-Stoer method \citep{bulirsch64} to integrate the orbit of `Oumuamua with a time step of 1\,kyr under perturbations from the Sun and the Galactic tide back to 100\,Myr ago. According to our tests, the energy and angular momentum are conserved to a precision of $10^{-8}$ over 1\, Gyr \citep{feng18a}. We further identify close encounters by comparing `Oumuamua's orbit with the orbits of the 0.23 million stars in the FS catalog \citep{feng17d}. Finally, we identify 109 encounters with periastron less than 5\,pc and with reliable astrometry and radial velocity data. By drawing 1000 clones from the uncertain initial conditions for each encounter and integrating their orbits, we calculate the encounter parameters and their uncertainties. We use 5\% and 95\% quantiles to measure the uncertainty. We use the minimum encounter distance, $d_{\rm enc}$ to represent the 5\% quantile since small $d_{\rm enc}$ is typically not well sampled (see \citealt{feng17d} for details). The results for the 109 encounters are available at \url{http://star-www.herts.ac.uk/~ffeng/Oumuamua/}.

From this sample, we select encounters either with periastron less than 2\,pc or with relative velocity less than 10\,km/s and show them in Table \ref{tab:sample}. These encounters are plausible candidates of origin because the probability of finding a random encounter with relative velocity less than 10\,km/s is about $7\times 10^{-3}$, assuming an Maxwell-Boltzmann distribution for the encounter velocity with a mean velocity of 53\,km/s \citep{rickman08,feng14}. For example, the probability of finding HIP 113020 with $v_{\rm enc}=4.91$\,km/s and $d_{\rm enc}=2.36$\,pc in the sample of 24 encounters with $d_{\rm enc}<2.5$\,pc is about 2\%. On the other hand, according to the conservation of energy, `Oumuamua would be significantly decelerated during its ejection, leading to a relative velocity typically less than 5\,km/s \citep{zuluaga17}. Thus the slow and close encounters in Table \ref{tab:sample} are rare but plausible candidates for the origin of `Oumuamua.

Since we only integrate orbits backward, the stars currently close to `Oumuamua have encounter time, $t_{\rm enc}=0$\,Myr. We find three encounters, HIP 104539, 17288, and 103749, with $d_{\rm enc}<2$\,pc and encounter velocity, $v_{\rm enc}<20$\,km/s. Among them, HIP 104539 is a A1V-type star with a mass of 2.70$\pm$0.58\,$M_\odot$ and age of 0.618$\pm$0.419\,Gyr \citep{zorec12}. However, its radial velocity is 12.0$\pm$4.4\,km/s \citep{gontcharov16}, leading to a large uncertainty in its encounter distance and velocity. Given the considerable plausibility of this candidate, further radial velocity is warranted to refine its trajectory. Another candidate HIP 17288 is an F5V-type binary with a mass of 1.2$\pm$0.1\,$M_\odot$ and an age of about 3.8\,Gyr \citep{david15}. HIP 103749 is also an F5-type binary with a total mass of about 2\,$M_\odot$ \citep{tokovinin14} and an age of about 3\,Gyr \citep{olsen84}. Our sample partly overlaps with the catalog provided by \cite{dybczynski17} but does not overlap with the one provided by \cite{zwart17}. Like \cite{dybczynski17} , we identify HIP 113020 (GJ 876) as a very slow encounter which passes `Oumuamua at 2.36\,pc. We also find fast encounters like HIP 3757 and HIP 3829 but they have very noisy spectra and hence poor quality radial velocities. Hence we do not report them in Table \ref{tab:sample} despite their small perihelia. In summary, the encounter parameters for the above candidates are still too uncertain to be confirmed as the origin of `Oumuamua partly because their encounter time is far in the past and their trajectories are not well constrained based on the current data. 

\begin{table*}
\caption{Selected sample of encounters of `Oumuamua and their nominal, 5\% and 95\% quantiles of $t_{\rm enc}$, $d_{\rm enc}$, and $v_{\rm enc}$. The encounters are sorted in an increasing order of $d_{\rm enc}^{\rm nom}$.}
\label{tab:sample}
\centering 
\begin{tabular}{lccccccccccc}
  \hline
Name&$t_{\rm enc}^{\rm nom}$& $t_{\rm enc}^{\rm 5\%}$& $t_{\rm enc}^{\rm 95\%}$ &$d_{\rm enc}^{\rm nom}$&$d_{\rm enc}^{\rm 5\%}$ &$d_{\rm enc}^{\rm 95\%}$ &$v_{\rm enc}^{\rm nom}$ &$v_{\rm enc}^{\rm 5\%}$ &$v_{\rm enc}^{\rm 95\%}$ \\
&(Myr)&(Myr)&(Myr)&(pc)&(pc)&(pc)&(km/s)&(km/s)&(km/s)\\
  \hline
HIP 21553 & -0.28 & -0.29 & -0.28 & 1.08 & 0.99 & 1.15 & 34.92 & 34.78 & 35.08 \\ 
  HIP 71681 & 0.00 & 0.00 & 0.00 & 1.26 & 1.13 & 1.33 & 36.23 & 34.44 & 38.04 \\ 
  HIP 70890 & 0.00 & 0.00 & 0.00 & 1.30 & 1.28 & 1.31 & 37.23 & 36.54 & 37.92 \\ 
  HIP 71683 & 0.00 & 0.00 & 0.00 & 1.33 & 1.30 & 1.34 & 35.28 & 34.24 & 36.31 \\ 
  HIP 17288 & -6.79 & -7.20 & -6.38 & 1.34 & 0.07 & 7.81 & 14.85 & 14.38 & 15.37 \\ 
  HIP 104539 & -10.51 & -25.14 & -5.80 & 1.42 & 0.25 & 54.72 & 10.20 & 3.28 & 17.16 \\ 
  TYC 7582-1449-1 & -8.97 & -9.87 & -8.00 & 1.55 & 0.68 & 26.34 & 22.11 & 20.86 & 23.67 \\ 
  HIP 101180 &-0.24 & -0.24 & -0.23 & 1.67 & 1.61 & 1.70 & 32.75 & 32.58 & 32.91 \\ 
  HIP 24608 &  -0.49 & -0.50 & -0.49 & 1.75 & 1.63 & 1.82 & 25.87 & 25.76 & 25.99 \\ 
  HIP 86916 &  -0.46 & -0.53 & -0.40 & 1.78 & 1.39 & 2.09 & 43.43 & 37.05 & 49.46 \\ 
  HIP 87937 &  0.00 & 0.00 & 0.00 & 1.81 & 1.80 & 1.82 & 134.91 & 134.57 & 135.24 \\ 
  TYC 5855-2215-1 & -6.65 & -7.70 & -5.51 & 1.93 & 0.51 & 72.38 & 40.23 & 38.65 & 43.61 \\ 
  HIP 103749 & -4.37 & -4.71 & -4.03 & 1.95 & 0.05 & 6.21 & 12.07 & 11.43 & 12.74 \\ 
  HIP 113020 & -0.81 & -0.83 & -0.78 & 2.36 & 2.14 & 2.50 & 4.91 & 4.65 & 5.17 \\ 
  HIP 107556 & -1.57 & -1.89 & -1.34 & 3.31 & 2.38 & 4.03 & 7.14 & 5.82 & 8.40 \\ 
  HIP 37766 & -0.57 & -0.59 & -0.55 & 3.57 & 3.22 & 3.78 & 8.20 & 7.80 & 8.61 \\ 
  HIP 51966 & -5.02 & -6.00 & -4.16 & 4.51 & 0.33 & 15.30 & 7.88 & 6.60 & 9.48 \\
  \hline
\end{tabular}
\end{table*}

\begin{figure}
\centering 
  \includegraphics[scale=0.5]{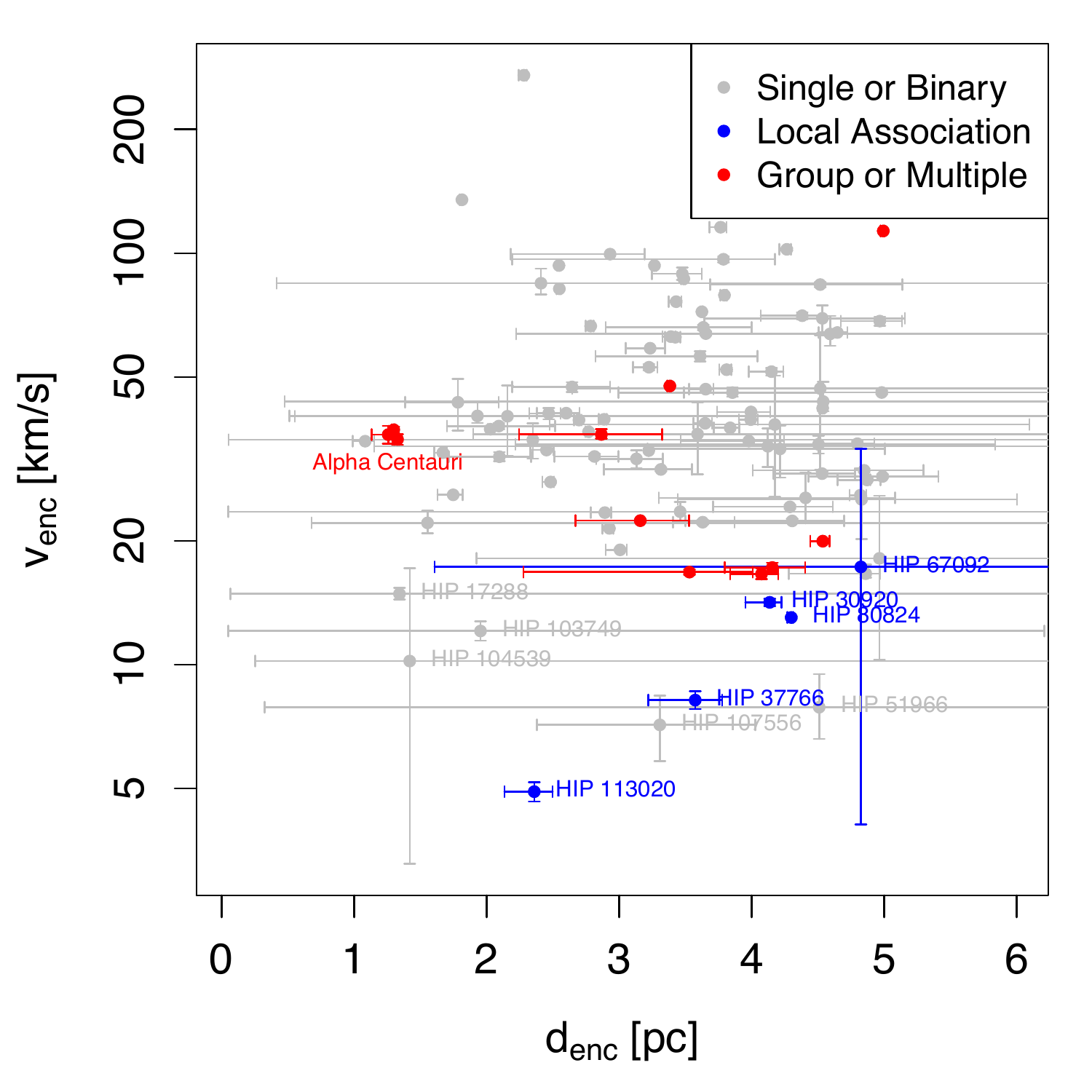}
  \caption{Distribution of $d_{\rm enc}$ and $v_{\rm enc}$ for the 109 encounters of `Oumuamua. The names of some interesting encounters are labeled on the right side of or below corresponding points.}
  \label{fig:encounter}
\end{figure}

We also investigate the origin of `Oumuamua by investigating its connection with nearby stellar groups and associations. We show the distribution $d_{\rm enc}$ and $v_{\rm enc}$ for different types of encounters in Fig. \ref{fig:encounter}. There are five encounters belonging to the Local Association (or Pleiades Moving Group), including stars associated with Pleiades, $\alpha$ Per, NGC 2516, IC 2602, and Scorpius-Centaurus \citep{eggen75, eggen95}. On the other hand, there are only six fast encounters belonging to five other moving groups and associations. According to \cite{montes01}, the Local Association has an age ranging from 20 to 150\,Myr and a mean $UVW$ of $(-11.6, -21.0, -11.4)$\,km/s which only differs from the `Oumuamua's veloctiy by 4\,km/s. Although the Carina and Columba Associations do have similar velocities \citep{gaidos17} and many of these group members are included in the FS catalog, we find no encounters belonging to these two groups. Considering that some members of the Local Association have approached `Oumuamu with small relative velocity and distance, `Oumuamua was probably ejected from a young stellar system in the Local Association. We further constrain its origin and age in the following section.  

\section{`Oumuamua is probably young}\label{sec:age}

\subsection{Kinematic constraint}\label{sec:dynamic}
The population of asteroids and comets is depleted by the accretion and scattering process during the formation of planets. Thus `Oumuamua is likely to be ejected from a stellar system during its early evolution when the system is dynamically hot. Hence the age of `Oumuamua is approximately the time scale of its migration in the Galaxy after being ejected. As observed by \cite{gaidos17}, `Oumuamua moves relatively slowly with respect to the Local Standard of Rest (LSR). The velocity difference is less than 10 and 3\,km/s for the LSR determined by \citealt{coskunoglu11} and by \citealt{schoenrich10}, respectively. Such a low velocity difference is also observed in many young stellar associations \citep{montes01,torres08}, and thus seems to support a young age of `Oumuamua. As an interstellar object migrates in the Galaxy, its dynamics would potentially be altered by stars, molecular clouds, spiral arms, star clusters, etc. and gradually deviate from the LSR. For example, Oumuamua's encounter with the Solar System will significantly alter its orbit and drive it away from the LSR. This so-called ``disk heating'' mechanism is intensively studied and observed (e.g., \citealt{dehnen98,holmberg09}). For example, the total velocity dispersion increases from $\sim$30\,km/s to $\sim$60\,km/s if the age $\tau$ (in units of Gyr) increases from 1 to 10\,Gyr following the relation of $\sigma_{\rm tot}\sim \tau^{0.34}$ according to \cite{holmberg09}. Assuming a similar heating mechanism for `Oumuamua-like objects, the probability of observing them with a velocity less than 10\,km/s with respect to the LSR would be 0.50, 0.26, and 0.13 for an age of 0.1\,Gyr, 1\,Gyr and 10\,Gyr, respectively. This probability would be halved for a peculiar velocity less than 5\,km/s (e.g., with respect to the LSR determined by \citealt{schoenrich10}). Moreover, low mass objects are more likely to be scattered by encounters according to the conservation of momentum. Thus encounters will change the orbits of low mass objects more significantly, which is one of the reasons why low mass (or late type) stars tend to have higher velocity dispersion than massive ones (e.g., fig. 5 of \cite{dehnen98}). Therefore, the kinematics of `Oumuamua favors a recent origin or ejection. 

\subsection{Physical constraint}\label{sec:rotation}
\begin{figure*}[t]
  \centering
  \vspace{-0.2in}
\includegraphics[scale=0.5]{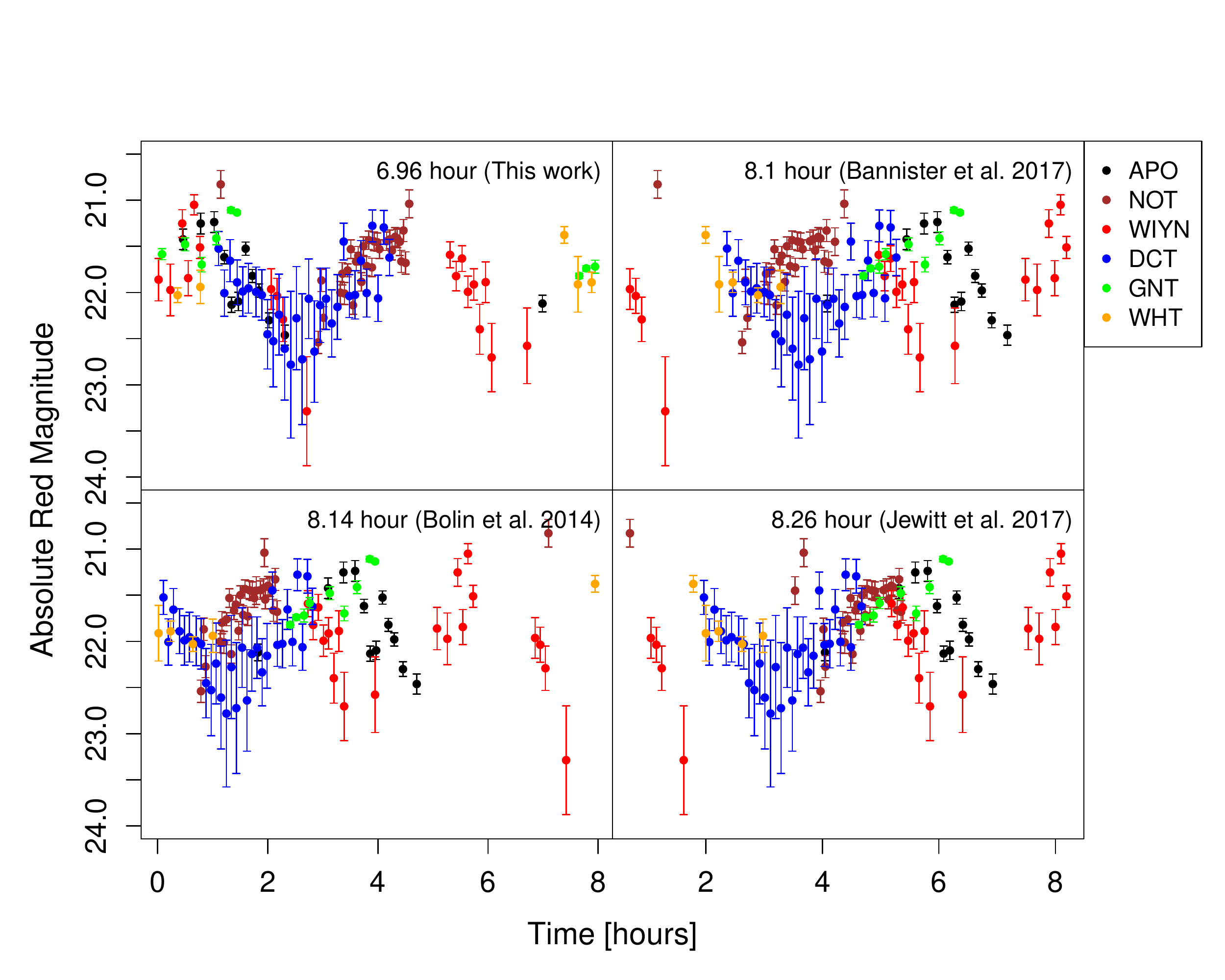}
\vspace{0.1in}
  \caption{Phased absolute R magnitude of `Oumuamua measured by various telescopes. The rotation periods determined in this work and in previous works are shown in the top right corner of panels. }
  \label{fig:fit}
\end{figure*}

We also investigate the origin of `Oumuamua by estimating its rotation period and axis ratio using the light curves measured by the Nordic Optical Telescope (NOT) and the Wisconsin-Indiana-Missouri-NOAO telescope (WIYN) \citep{jewitt17}, by the Apache Point Observatory (APO) \citep{bolin17}, by the Discovery Channel Telescope \citep{knight17}, and by the Frederick C. Gillett Gemini North Telescope (GNT) and the William Herschel Telescope (WHT) \citep{bannister17}. We convert different magnitudes into R magnitude and calculate the absolute magnitude using equation 1 of \cite{jewitt17} and the phase angles from \cite{jewitt17} and \cite{knight17}. We use a sinusoidal function in combination with the first order moving average noise model in \cite{feng17c} to estimate the rotation period and magnitude variation in the Bayesian framework introduced in \cite{feng16}. We find a double peaked rotation period of $6.96_{-0.39}^{+1.45}$\,h. The magnitude variation is about 2.0$\pm$0.2 mags, corresponding to an axis ratio of about 6:1 and semi-axes of 230\,m$\times$35\,m if `Oumuamua is a prolate ellipsoid \citep{jewitt17}. The phased red magnitudes subtracted by a best-fit linear trend for different rotation periods determined in the literature is shown in Fig. \ref{fig:fit}. By using all available data sets, we identify a shorter rotation period, compared with previous values. It is evident that the NOT and WIYN data are not well modeled by the phase curve for 8.1\,h estimated from DCT, GNT and WHT by \cite{bannister17}. \cite{bolin17} estimate a rotation period of 8.14\,h based only on the APO and DCT data, leading to a poor modeling of other data sets. Although \cite{jewitt17}'s model well fit the DCT, NOT and WIYN data, they poorly fit the other data sets. However, our estimation of a rotation period of 6.96\,d is favored by all data sets despite being considerably different than previous estimation. We note the work by \cite{meech17} who estimate a similar rotation period of 7.34$\pm$0.06\,h based on a total of 131 observations from the Very Large Telescope (VLT), Keck 2 Telescope, Canada-France-Hawaii Telescope (CFHT), United Kingdom Infrared Telescope (UKIRT), and the Gemini South Telescope. 

 According to \cite{meech17}, `Oumuamua is a red and extremely elongated interstellar asteroid with an axis ratio of 10:1 if modeling the light curve with a triaxial ellipsoid. A combination of the data from \cite{meech17} and the other data sets may lead to an axis ratio between 6:1 to 10:1. Such an elongated shape is rarely seen in the Solar System. Its neutral or slightly red color \citep{bannister17,meech17} indicates an organic-rich surface found in comets/asteroids in the outer Solar System although no cometary activity has been detected \citep{ye17}. Considering that the cometary population is a few orders of magnitude higher than the asteroid population in well-evolved Sun-like systems \citep{feng15b}, `Oumuamua was more likely to be ejected from the middle part of a young stellar system. If it was too close to the star, it is unlikely to be organic rich. But if it was too far away from its host star, it would be icy and show cometary activity during its encounter with the Solar System. Moreover, a young stellar system is dynamically hot, and abundant in debris objects, and thus are more likely to be the source of interstellar objects like `Oumuamua. On the other hand, its extremely elongated shape and high density \citep{meech17} is probably related to energetic collisions between minor bodies or planets such as the Late Heavy Bombardments caused by planet migration \citep{gomes05}. In addition, the color of `Oumuamua is not as red as some Kuiper Belt Objects (KBOs) which have been reddened by space weathering such as cosmic ray and interstellar medium \citep{jewitt02,jedicke04}. Hence it seems less likely to have travelled for Gyrs before encountering the Solar System. 

 It is interesting to consider the density of interstellar star formation debris, based on \cite{zwart17} and \cite{hanse17}'s numerical investigations, the population of unbound non-cometary asteroidal objects is much larger than that of cometary objects. Thus along with the likelihood of such objects being readily scattered to a higher velocity distribution (Section \ref{sec:dynamic}) it appears that they are an unaccounted for constituent of the mass of the halo of our galaxy. 

To derive the density of `Oumuamua-like objects, we use the following equation to model the encounter rate $F$, 
\begin{equation}
  F=n\sigma v_{\rm enc}~,
\end{equation}
where $n$ is the number density of interstellar objects, $v_{\rm enc}$ is encounter velocity, $\sigma$ is its cross section. The cross section is approximately $\pi d_{\rm max}^2$, where $d_{\rm max}$ is the maximum encounter distance. The mean encounter velocity is at least 50\,km/s according to \cite{rickman08,feng14, feng17d}. Since `Oumuamua is the first interstellar object humans have so far recognised, we assume that the encounter rate of an `Oumuamua-like object (with a size $\gtrsim$100\,m) with impact parameter less than 0.5\,au\footnote{This roughly correspond to a perihelion of $\sim$0.3\,au for a interstellar object with $v_{\rm enc}\sim50$\,km/s and a maximum apparent magnitude of $\sim$20\,mag for an `Oumuamua-like object.} per 20 years\footnote{The asteroid surveys of Pan-STARRS1 \citep{kaiser10}, Catalina Sky Survey, and the Mt. Lemmon Survey \citep{christensen12} find no interstellar objects in the past 19 years \citep{engelhardt17}.} is 1. Hence there would be $1.4\times 10^{13}$ interstellar objects with mean diameter larger than 100\,m per pc$^{3}$ or $6.0\times10^{-3}$\,au$^{-3}$, which is higher than the value of $1.4\times10^{-4}$\,au$^{-3}$ derived by \cite{engelhardt17} who consider interstellar objects with $>1$\,km diameter.

Our value is lower than the density derived by \cite{zwart17} since they only consider the non-detection in the Pan-STARRS1 survey. It is also slightly lower than that in \cite{laughlin17} probably because they have adopted a low mean encounter velocity. Since the sensitivity of asteroid surveys to `Oumuamua-like objects increases with time, the non-detection period could be shorter than 20 years. Hence our estimation is a lower limit of the abundance of interstellar objects.
\section{Discussion and conclusion}\label{sec:conclusion}
Based on the kinematics of `Oumuamua, we find 109 encounters with a nominal encounter distance less than 5\,pc. There are 17 stars with an encounter distance less than 2\,pc or with relative velocity less than 10\,km/s. Five slow encounters in the whole sample belong to the Local Association while most of the others are field stars, indicating an origin of `Oumuamua in the Local Association. We note that the reader might be wondering about a future observer in some other solar system who might detect `Oumuamua and integrate its orbit backwards to discover that the object came directly from the Solar System and then conclude a Solar System origin. While this is a possible way to underestimate the age of `Oumuamua, we argue that velocity is more important than distance in finding candidates since velocity follows an Maxwell distribution while distance follows a power law distribution. We find that slow and close encounters are rare but plausible candidates for the origin of `Oumuamua.

Moreover, we consider that `Oumuamua's low velocity with respect to the LSR indicates a short period of interstellar travel. The interpretation of `Oumuamua having a relatively young age is further supported by its relatively neutral color due to a lack of long-term exposure to bombardments from the interstellar medium and cosmic rays. Its extremely elongated shape is rarely seen in the Solar System and is probably caused by energetic events such as planetary collisions and impacts. It is asteroidal and its surface is organic rich but without observable cometary activities, suggestive of an origin in the middle part of a young stellar system.

We estimate a number density of at least $6.0\times10^{-3}$\,au$^{-3}$ for interstellar objects with diameter larger than 100\,m, in agreement with previous results. Such a number density seems to be much lower than the expected value assuming that extra-solar systems form in a similar way as the Solar System \citep{engelhardt17}. This discrepancy is probably not due to a different formation mechanism as \cite{engelhardt17} suggest but due to the gravitational scattering of interstellar objects by stars and floating planets. According to the conservation of momentum, low-mass objects are more likely to be scattered than high-mass ones and thus such objects more easily accelerated to escape the Galaxy or to float into the Galactic halo.

Current microlensing surveys are sensitive down to objects with masses of so called super-Earth planets \citep{mroz17}. Future missions such as WFIRST are expected to probe masses down to that of Mars \citep{spergel15}. Nonetheless more objects such as Oumuamua will enable a local determination of the density of unbound debris from star formation and thus a comparison with expected interstellar planetesimal flux from the star formation process and an estimation of the contribution of such objects to the mass of the Galactic halo.

Interstellar objects may also bombard the Earth and cause catastrophic events such as mass extinctions \citep{bailer-jones09}. Since these objects are anisotropic in velocity due to the solar apex motion \citep{feng14}, they would probably form anisotropic impact craters on terrestrial planets and moons such as the lunar craters \citep{williams16}. The high velocity of interstellar objects means that for a given size and frequency they have the potential to cause relatively more catastrophic events such as mass extinctions \citep{alvarez84} than Solar System minor bodies.

Our search of the origin home of `Oumuamua is limited by the precision of astrometry and radial velocity data. The upcoming Gaia data releases \citep{brown17} will provide accurate astrometry and stellar parameters for more stars and thus enable a more comprehensive study for the origin of `Oumuamua. 

\section*{Acknowledgements}
FF and HJ are supported by the Science and Technology Facilities Council (ST/M001008/1). Encounters are identified using the data partly provided by the databases operated at CDS, Strasbourg, France and calculations of orbital parameters using the JPL's HORIZONS system, and data obtained from the the International Astronomical Union's Minor Planet Center. We are very grateful for the insightful comments of the anonymous referee.
\bibliographystyle{aasjournal}
\bibliography{dynamics}
\end{document}